\documentclass{INTERSPEECH2023}

\usepackage{amsmath,graphicx}

\usepackage{caption} 
\usepackage{tabularx}
\usepackage{float}
\usepackage{array}  
\usepackage{booktabs} 
\usepackage{multirow}
\usepackage{makecell}

\usepackage{color}

\usepackage{bbding}

\usepackage{amsfonts,amssymb} 
\usepackage{bbding}  
\usepackage{caption} 
\usepackage{tabularx} 
\usepackage{float}
\usepackage{array}  
\usepackage{booktabs} 
\usepackage{multirow}
\usepackage{makecell} 
\usepackage{color}
\usepackage{todonotes}

\usepackage{subfigure} 

\interspeechcameraready

\title{Joint Prediction of Audio Event and Annoyance Rating in an Urban Soundscape by Hierarchical Graph Representation Learning}

\name{Yuanbo Hou$^1$, Siyang Song$^2$, Cheng Luo$^3$,  Andrew Mitchell$^4$, Qiaoqiao Ren$^5$, \\ Weicheng Xie$^3$, Jian Kang$^4$, Wenwu Wang$^6$, Dick Botteldooren$^1$}
\address{
  $^1$WAVES Research Group, Ghent University, Belgium. 
  $^2$University of Leicester, UK. \\
  $^3$Shenzhen University, China.
  $^4$University College London, UK. \\
  $^5$AIRO-IDlab, Ghent University-Imec, Belgium. 
  $^6$University of Surrey, UK.
  }
\email{\{Yuanbo.Hou, Dick.Botteldooren\}@UGent.be}

\begin{document}

\maketitle
 
\begin{abstract} 
\noindent Sound events in daily life carry rich information about the objective world. The composition of these sounds affects the mood of people in a soundscape. Most previous approaches only focus on classifying and detecting audio events and scenes, but may ignore their perceptual quality that may impact humans' listening mood for the environment, e.g. annoyance. To this end, this paper proposes a novel hierarchical graph representation learning (HGRL) approach which links objective audio events (AE) with subjective annoyance ratings (AR) of the soundscape perceived by humans. The hierarchical graph consists of fine-grained event (fAE) embeddings with single-class event semantics, coarse-grained event (cAE) embeddings with multi-class event semantics, and AR embeddings. Experiments show the proposed HGRL successfully integrates AE with AR for AEC and ARP tasks, while coordinating the relations between cAE and fAE and further aligning the two different grains of AE information with the AR.

\end{abstract}

\section{Introduction}\label{section1}

Audio event classification (AEC) aims to recognise predefined semantic events from audio clips to indicate whether the audio clip contains target events. 
AEC is useful for acoustic scene recognition \cite{scene}, robot hearing \cite{ren2016sound}, and monitoring \cite{monitor}.
Meanwhile, human annoyance rating prediction (ARP) aims to predict the annoyance rating (AR) given by humans to express their appraisal of a soundscape containing multiple undesired audio events.
Joint APR and AEC can be used in soundscape design \cite{soundscape}, human-robot interaction \cite{hri_audio}, and smart cities \cite{city}. While AEC focuses on describing which audio events (AE) are present in the soundscape, ARP aims to identify how the combination of these sound events may induce the particular listening mood, i.e. annoyance, for humans in the soundscape. 
While noise annoyance at the community level has typically been explored in reference to coarse-grained sound classes, e.g. traffic and aircraft noise \cite{Kryter1994Handbook}, soundscape pleasantness (considered the opposite of annoyance) more often considers specific audio events such as bird sounds \cite{Hao2016Assessment} (occasionally even differentiating between different species of birds sounds \cite{Ratcliffe2013Bird}) or water fountain sounds \cite{Trudeau2020Tale}. Psychoacoustic annoyance is also defined at a coarse-grained level, originally formulated to be applied to consumer products such as vacuums, refrigerators, and car engines \cite{Zwicker2007Psychoacoustics}. 
As such, both coarse-grained audio events (cAE) and fine-grained audio events (fAE) are considered in this paper.

To capture the high-level acoustic representations of AE, convolutional neural networks (CNN)-based models with local receptive fields have been widely proposed for AEC tasks, which show outstanding performance \cite{kong2020panns}. 
The convolutional recurrent neural networks (CRNN) \cite{audio_tagging, wei2020crnn} combining recurrent layers, which excel in temporal modelling, further enhance the model's ability to recognize diverse AE. 
The Audio Spectrogram Transformer \cite{ast} has also been proposed for AEC, which outperforms the CNN-based pretrained audio neural networks (PANNs) \cite{kong2020panns} on the large-scale audio event dataset AudioSet \cite{audioset}, thanks to the ability of the Transformer \cite{Transformer} with multi-headed attention (MHA) for modelling long-term dependency information in audio clips.
However, global attention in MHA may smooth out the boundaries between audio events and background noises \cite{edc}. To alleviate this problem, event-related data conditioning \cite{edc} is proposed for AEC. 
Audio events in real-life audio clips are usually not isolated, but exist as temporal sequences. To exploit this property, CRNN-based models \cite{wangyun} and contextual Transformer \cite{ctsat} have been proposed for sequential audio tagging.

The above studies aim to provide an objective description of the content of audio clips by exploring what audio events are present.  
In real life, various audio events in soundscapes can be overlapped and coupled to form polyphonic audio clips, bringing people different perceptual experiences and listening moods \cite{Oldoni2013computational, delta}.
For example, people can feel relaxed and pleasant when hearing the sound of running water and bird songs in a park scene, while they can be annoyed when hearing the noise of speeding cars and harsh horns on the street scene. Previous studies on soundscape and its impact on people, have acknowledged the importance of the source of sound but did not elaborate on matching indicators. Rather, they focused on noise levels, psycho-acoustic indicators such as calculated loudness, and various other indicators for the overall sound \cite{Lionello2020systematic, Mitchell2021Investigating}.
This paper aims to combine AEC with human perception-related ARP. 
Furthermore, this paper explores the feasibility of predicting AR, which is a dominant performance metric for perceptual evaluation of soundscapes, based on the detected AE.

\label{ssec:figure-f}
\begin{figure*}[t] 
	\setlength{\belowcaptionskip}{-0.6cm}   
	\centerline{\includegraphics[width = 0.95 \textwidth]{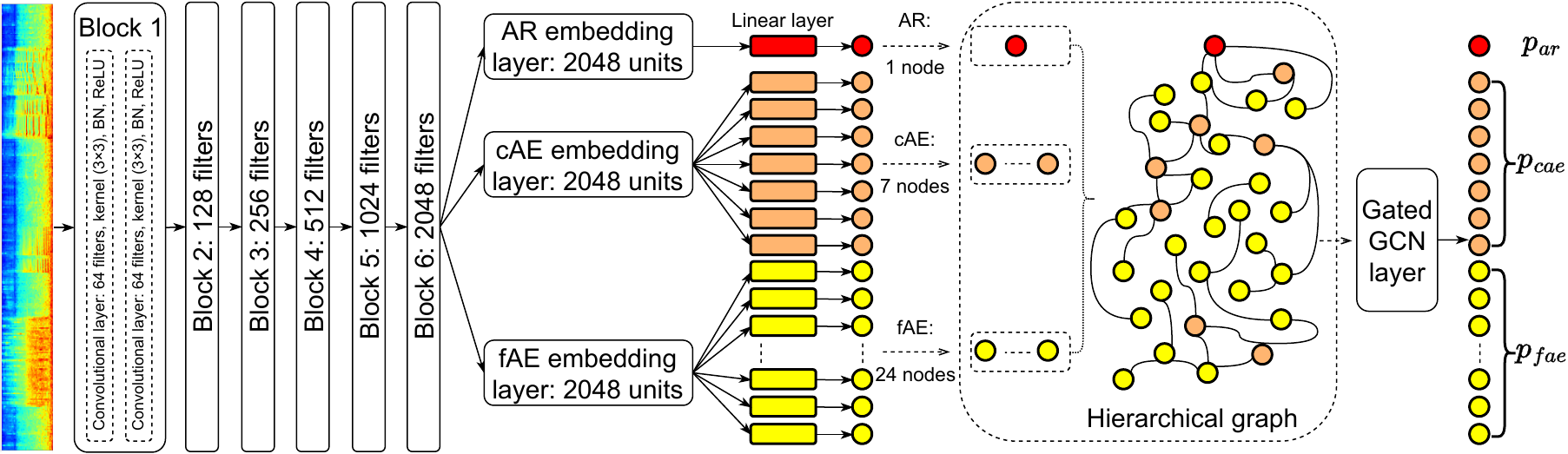}}
	\caption{The proposed hierarchical graph representation learning (HGRL) with fAE-cAE-AR (fcAR) graph.}
	\label{model}
\end{figure*}

Inspired by the scene-dependent event relational graph representation method \cite{graph}, this paper proposes a hierarchical graph representation learning (HGRL) method, which performs AEC and ARP tasks simultaneously, and is trained on the public DeLTA dataset \cite{delta} with 24-class AE labels and the overall human AR.  
To enrich the semantic information of the hierarchical graph learned by HGRL, we summarise the 24-class fAE into 7-class cAE referred to the label topology in AudioSet \cite{audioset}. 
Next, a three-level hierarchical graph representation, which consists of the low-level fAE graph, mid-level cAE graph, and top-level human AR graph, is defined for each audio clip.
The hierarchical graph with different semantic nodes and edges is input to a gated graph convolutional network (Gated GCN) \cite{gated_GCN} to classify 24-class fAE and 7-class cAE, and to predict AR. The fAE labels and AR scores are inherent in the DeLTA dataset, and the cAE labels are defined in Section \ref{dataset}.

The contributions are summarized as follows. 1) This paper links AEC with ARP related to human appraisal of a soundscape; 
2) Inspired by the relations between fAE and human AR, this paper proposes the HGRL that models the relations of AE and their connections with AR; 
3) Leveraging the knowledge from human perception and appraisal of soundscapes, this paper proposes cAE classes based on fAE and embeds them into the hierarchical graph to serve as an information exchange layer between fAE and AR; 
4) This paper intuitively visualizes the learned relations between the fAE, cAE classes and AR to illustrate the ability of the hierarchical graph model to coordinate and align different levels of audio information.

\vspace{-0.2cm}
\section{Hierarchical graph representation learning for AEC and ARP}

This section describes the dataset, the strategy of node feature extraction for AE and AR, based on which a hierarchical graph representation can be learned to perform AEC and ARP tasks.

\vspace{-0.1cm}
\subsection{Dataset}\label{dataset}
\vspace{-0.1cm} 
To the best of our knowledge, DeLTA \cite{delta} is the only publicly available dataset that includes both AE labels and human AR scores. 
Each audio clip in DeLTA has a clip-level 24-dimensional multi-hot vector as the fAE label, and an AR (continuously from 1 to 10). 
DeLTA comprises 2890 15-second binaural audio clips, where the training, validation, and test sets contain 2200, 245, and 445  audio clips, respectively.
Based on the ontology of the event labels in AudioSet \cite{audioset}, we further group the 24 fAE classes of DeLTA into 7 cAE classes: 1) \textbf{\textit{Vehicle}}: [\textit{aircraft}, \textit{bus}, \textit{car}, \textit{general traffic}, \textit{motorcycle}, \textit{rail}, \textit{screeching brakes}];
2) \textbf{\textit{Music}}: [\textit{bells}, \textit{music}];
3) \textbf{\textit{Animals}}: [\textit{bird tweet}, \textit{dog bark}];
4) \textbf{\textit{Human sounds}}: [\textit{children}, \textit{laughter}, \textit{speech}, \textit{shouting}, \textit{footsteps}];
5) \textbf{\textit{Alarm}}: [\textit{siren}, \textit{horn}];
6) \textbf{\textit{Natural sounds}}: [\textit{rustling leaves}, \textit{water}];
7) \textbf{\textit{Other}}: [\textit{construction}, \textit{non-identifiable}, \textit{ventilation}, \textit{other}].

\vspace{-0.1cm}
\subsection{Semantic node feature extraction}
\vspace{-0.1cm}
Given an input audio clip represented in the time-domain, we first converted it into a time-frequency domain spectrogram, which is fed into convolutional blocks to extract audio representations suitable for the AEC and ARP tasks.

As illustrated in Figure~\ref{model}, the employed convolutional block refers to the convolutional block structure in PANNs \cite{kong2020panns}. Each convolutional block uses a VGG-like CNN \cite{vgg}. That is, the convolutional layer is repeated twice and followed by batch normalization (BN) \cite{batchnormal}, and ReLU activation functions \cite{relu}.
Unlike PANN, the model proposed in Figure~\ref{model} connects 3 2048-unit embedding layers in parallel after convolutional blocks to convert high-level audio representations into separate semantic embeddings, respectively. Based on these embeddings, we can obtain the features of different nodes in the graph.

\textbf{AR node.}
The AR embedding layer outputs a rating embedding of dimension (B, 2048), where B is the batch size. The rating embedding is mapped to an advanced rating representation by a linear layer with 64 units. Since the AR graph contains only one node, the AR representation (B, 64) is deformed to (B, 1, 64) and used as the AR node feature.

\textbf{cAE nodes.}
To independently learn the node features of each type of cAE, the cAE embeddings with the dimension of (B, 2048) are respectively input into 7 64-unit linear layers. 
Among them, the output of each layer is (B, 64). After concatenating the outputs of 7 layers, a tensor with a dimension of (B, 7, 64) is obtained as the feature of the cAE nodes.

\textbf{fAE nodes.}
The process for extracting the feature of the fAE nodes is similar to the extraction of the feature of the cAE nodes. 
The difference is that the fAE has 24 classes of AE, so the fAE embeddings are separately fed into 24 64-unit linear layers to capture the representation of each class of AE. 
Finally, the outputs of the 24 linear layers are concatenated into a (B, 24, 64) tensor as the feature of the fAE nodes.

\vspace{-0.1cm}
\subsection{Construction of hierarchical graphs}\label{graph}
\vspace{-0.1cm}
 
This paper proposes to learn a novel hierarchical graph which consists of two types of graphs as discussed next.

\vspace{-0.1cm}
\subsubsection{The fAE-AR (fAR) graph}\label{far}
\vspace{-0.1cm}

As mentioned in Section \ref{dataset}, the DeLTA dataset \cite{delta} used in this paper does not contain labels on cAE. 
Therefore, an intuitive idea is to construct a hierarchical graph composed of fAE and AR directly. 
The fAE-AR (fAR) graph does not contain the cAE embedding layer and the corresponding cAE nodes shown in Figure~\ref{model}, so there is no prediction vector $p_{cae}$ of cAE either. For the fAR-based HGRL, the final loss is  
\begin{equation}
\setlength{\abovedisplayskip}{3pt}
\setlength{\belowdisplayskip}{3pt} 
\mathcal{L}_\text{fAR} = \mathcal{L}_\text{fAE} + \mathcal{L}_\text{AR}
\end{equation}
where $\mathcal{L}_\text{fAE}$ is the fine-grained AE classification loss and $\mathcal{L}_\text{AR}$ is the AR regression loss.
Given ${y}_{fae}$ and ${y}_{ar}$ as the true labels for fAE and AR, respectively, binary cross entropy (BCE) is used as the loss function in the fine-grained AEC, $\mathcal{L}_\text{fAE} = BCE(p_{fae}, y_{fae})$, and 
mean squared error (MSE) loss is used as the loss function in the ARP, 
$\mathcal{L}_\text{AR} = MSE(p_{ar}, y_{ar})$.

\subsubsection{The fAE-cAE-AR (fcAR) graph}

Based on DeLTA's 24 fAE labels, this paper proposes 7 cAE labels in Section~\ref{dataset} and assigns them to each audio clip.
 The cAE is the summary of fAE.  
Compared with a wide variety of specific fAE, in a soundscape, people tend to perceive greater differences between sounds belonging to different cAE classes. 
Therefore, the cAE information can be used as an intermediate between the fAE information and the overall human perception (i.e. mood), to bridge the relations between many low-level fAEs and the top-level AR.

In the fcAR-based HGRL shown in Figure~\ref{model}, we expect to investigate whether the graph-based model can learn the information about cAE nodes without the supervised labels for cAE-related output $p_{cae}$, and improve the performance of the model on AEC and ARP tasks.
Therefore, for the fcAR-based HGRL in the unsupervised learning (UL) mode, the total loss is 
\begin{equation}
\setlength{\abovedisplayskip}{3pt}
\setlength{\belowdisplayskip}{3pt} 
\mathcal{L}_\text{fcAR-UL} = \mathcal{L}_\text{fAE} + \mathcal{L}_\text{AR}
\end{equation}
For the fcAR-based HGRL in supervised learning (SL) mode, the cAE nodes are trained with supervision, and the total loss is 
\begin{equation}
\setlength{\abovedisplayskip}{3pt}
\setlength{\belowdisplayskip}{3pt} 
\mathcal{L}_\text{fcAR-SL} = \mathcal{L}_\text{fAE} + \mathcal{L}_\text{AR} + \mathcal{L}_\text{cAE}
\end{equation}
where  $\mathcal{L}_\text{cAE} = BCE(p_{cae}, y_{ce})$, and ${y}_{ce}$ is true label for cAE.

The hierarchical graph, consisting of semantic nodes and edges, will be fed into the Gated GCN layer to further learn and update the features from the nodes and the corresponding edges by considering the information of the whole graph. 
The nodes features of (B, N, 64) output from the Gated GCN layer are pooled into (B, N), where N is the number of nodes in the graph. 
Then different nodes are directly used for the corresponding AEC and ARP tasks. Taking the fcAR-SL graph as an example, there are 32 nodes, so the final output dimension is (B, 32), the dimension of $p_{ar}$, $p_{cae}$, $p_{fae}$ are (B, 1), (B, 7), (B, 24), respectively, and these outputs are used directly in the corresponding loss functions.

\vspace{-0.2cm}
\section{Experiments and results}

\subsection{Baseline, Experiments Setup, and Metrics}

As this is the first paper performing AEC and human-related ARP regression tasks simultaneously, no reference models are available in the literature. 
Hence, the commonly used deep neural network (DNN), CNN, and CNN-Transformer are employed as baselines for comparison: 
\textbf{1)} the DNN consists of 4 fully connected (FC) layers followed by a ReLU function, the AEC and ARP layers. The number of units in each FC layer is 64, 128, 256 and 512, respectively. The output of the final FC layer is flattened and fed to the AEC and ARP layers, respectively. 
\textbf{2)} the CNN consists of 4 convolutional layers with (3 × 3) kernels, the AEC and ARP layers. The filters in each convolutional layer are 64, 128, 256 and 512, respectively. The output of the final convolutional layer is flattened and fed to the AEC and ARP layers, respectively. 
\textbf{3)} the CNN-Transformer consists of 3 convolutional layers with (3 × 3) kernels, a Transformer encoder \cite{Transformer}, the AEC and ARP layers.
Please check the project homepage 
{\footnotesize{(\textcolor{blue}{\underline{https://github.com/Yuanbo2020/HGRL}})}} 
for more details.

The log-mel energy with 64 banks \cite{mel} is employed as the acoustic feature, which is extracted by the Short-Time Fourier Transform (STFT) with a Hamming window length of 46 \textit{ms} and a window overlap of $1/3$ \cite{scene}. 
Dropout and normalization are used in training to prevent over-fitting of the model \cite{dropout}. 
A batch size of 64 and Adam optimizer \cite{adam} with a learning rate of 1e-3 are used to minimize the loss. 
The systems are trained on a card Tesla V100 GPU for 100 epochs.
Accuracy (\textit{Acc}), \textit{F-score} \cite{metrics}, and threshold-free area under curve (\textit{AUC}) \cite{AUC} are used to evaluate the classification results.
Mean absolute error (\textit{MAE}), mean square error (\textit{MSE}) and R2-score (\textit{R2}) \cite{r2} are used to measure the regression results.
Higher \textit{Acc}, \textit{F-score}, \textit{AUC}, \textit{R2} and lower \textit{MSE}, \textit{MAE} indicate better performance.

\vspace{-0.1cm}
\subsection{Results and Analysis}

\vspace{-0.1cm}
\textbf{Ablation study on the AE information and AR information.}
The proposed HGRL involves two types of foundational information: AE and AR. 
The fAE graph described in Section~\ref{far} only uses fAE and AR information. 
To investigate the impact of the individual and joint effects of the two types of information (AE and AR) on both AEC and ARP tasks, an ablation study is conducted as reported in Table~\ref{tab:ablation}.

\begin{table}[H]   
	\setlength{\abovecaptionskip}{0.2cm}   
	\setlength{\belowcaptionskip}{-0.2cm}  
	\renewcommand\tabcolsep{1pt} 
	\centering
	\caption{Ablation study of the fAR-based HGRL on the test set.}
	\begin{tabular}
	{p{0.2cm}<{\centering}|
	p{0.7cm}<{\centering}
	p{0.7cm}<{\centering}|
	p{1.2cm}<{\centering}
	p{1.55cm}<{\centering}
	p{1.cm}<{\centering}|
 p{1.cm}<{\centering}
 p{1.cm}<{\centering}}
	
		\toprule[1pt] 
		\specialrule{0em}{0.1pt}{0.1pt}

\multirow{2}{*}{\makecell[c]{\#}} & \multicolumn{2}{c|}{Information}	 &
\multicolumn{3}{c|}{AEC} & \multicolumn{2}{c}{ARP} \\
		\cline{2-8}     
	  &  \textit{fAE}  & \textit{AR}  & \textit{Acc.} (\%) & \textit{F-score} (\%) & \textit{AUC} & \textit{MSE} & \textit{MAE}  \\
	\hline 
		\specialrule{0em}{0.em}{0.pt}
		1 & \CheckmarkBold &  \XSolidBrush &  89.654 &  57.253 & 0.863 & \textit{N/A}  & \textit{N/A} \\ 

  2 & \XSolidBrush & \CheckmarkBold  &  \textit{N/A} &  \textit{N/A} & \textit{N/A} &  1.482  & 0.949 \\  

   3 & \CheckmarkBold & \CheckmarkBold  &  \textbf{90.895} & \textbf{59.781} & \textbf{0.874} & \textbf{ 1.364}  & \textbf{0.917} \\  

		\specialrule{0em}{0pt}{0em}
		\bottomrule[1pt]
	\end{tabular}
	\label{tab:ablation}
\end{table}

\vspace{-0.2cm}
In \#1 of Table~\ref{tab:ablation}, since only the AE information is used, the prediction of the AR information of the corresponding graph model is not available (N/A). 
Similarly, the AE information in \#2 is also N/A. 
Compared with \#1 and \#2, which use a single type of information, the graph-based HGRL using AE and AR information in \#3 improves the performance of AEC and ARP tasks, which shows that the proposed model can effectively combine AE information with AR information, and that jointly using the AE and AR information for the proposed model is beneficial for both tasks.


\textbf{Performance of different hierarchical graphs.}
According to the description in Section \ref{graph}, the hierarchical graph proposed can be subdivided into 3 types: 1) fAR composed of fAE and AR embeddings;  2) fcAR-UL consisting of fAE and AR embeddings, and unsupervised learning of cAE embeddings; 
3) fcAR-SL consisting of fAE and AR embeddings, and supervised learning of cAE embeddings. 
Table~\ref{tab:peformance} details the performance of these 3 types of hierarchical graph models on AEC and ARP  tasks. 
The convolutional layers in the feature extraction part refer to the convolutional layers in PANNs \cite{kong2020panns}, which are pre-trained on a large-scale audio dataset AudioSet \cite{audioset} with 527 classes of AE.  
In other words, the convolutional layer weights (ConW) in PANNs contain feature information from various AE. 
Therefore, an intuitive idea is whether introducing this diverse feature information into the feature extraction part of the proposed HGRL could improve the learning of the overall graph model.

In Table~\ref{tab:peformance}, the models that transfer the ConW from PANNs into HRGL's convolutional part (the rest of HGRL is randomly initialized in training) generally outperform those that do not. 
Therefore, using the ConW with richer representation extracted from AudioSet benefits learning HGRL.
When the models are randomly initialized without using ConW, the AEC accuracy of the fcAR-based model containing 24-class fAE and 7-class cAE is slightly inferior to that of the fAR-based model without 7-class cAE. 
The reasons may be 1) the supervised labels for the 7-class cAE in this paper are derived from the semantic topological map of AudioSet, and are inherently inaccurate; 
2) There are overlaps between the semantic labels of the 7-class cAE, which implicitly increase the difficulty of the model in identifying different cAE;
3) The feature extraction part of the proposed HGRL is inadequate for cAE composed of multiple classes of fAE, as evidenced by using ConvW for the models involving cAE in Table~\ref{tab:peformance}, where the AEC accuracy of both fAE and cAE is improved.
The fAR-based model does not contain cAE information, so its corresponding 7-class cAE classification accuracy is N/A. 
In fcAR-UL, the prediction of cAE $p_{cae}$ can be regarded as almost random, since there is no supervision and correction for the cAE-related output $p_{cae}$. Nevertheless, the performance of AR significantly improves.
Finally, fcAR-SL aided by ConW initialization performs best on the ARP task.

\begin{table}[H]   
	\setlength{\abovecaptionskip}{0.1cm}   
	\setlength{\belowcaptionskip}{-0.cm}  
	\renewcommand\tabcolsep{1pt} 
	\centering
	\caption{Performance of the proposed HGRL on the test set.}
	\begin{tabular}
	{p{1cm}<{\centering}|
	p{0.8cm}<{\centering} |
	p{0.8cm}<{\centering}|
	p{1.3cm}<{\centering} |
	p{0.9cm}<{\centering}|
 p{0.9cm}<{\centering}
 p{0.9cm}<{\centering}
 p{0.9cm}<{\centering}}
	
		\toprule[1pt]  
		\specialrule{0em}{0.1pt}{0.1pt} 

 PANNs & 
\multicolumn{2}{c|}{Hierarchical} &	\multicolumn{2}{c|}{AEC \textit{Acc.} (\%)} & \multicolumn{3}{c}{ARP} \\
		\cline{4-8}     
	 ConvW &  \multicolumn{2}{c|}{Graph}      & 24 fAE & 7 cAE & \textit{MSE} & \textit{MAE} & \textit{R2}  \\
	\hline 
		\specialrule{0em}{0.em}{0.pt}
		\multirow{3}{*}{\makecell[c]{\rotatebox{90}{Without}}} & \multicolumn{2}{c|}{fAR}  &  90.895 &  \textit{N/A} & 1.364  & 0.917 & 0.296 \\

    \cline{2-3} 
    & \multirow{2}{*}{\makecell[c]{\rotatebox{90}{fcAR}}}  & UL &  90.171 & 59.230 &  1.093  & 0.818 & 0.436  \\  
    
    \cline{3-3}   
    &  & SL &  90.459 &  82.009 &  1.076  & 0.817 & 0.444  \\ 
    
    \hline

    \multirow{3}{*}{\makecell[c]{\rotatebox{90}{With}}} & \multicolumn{2}{c|}{fAR}  &  91.751 &  \textit{N/A} & 1.176  & 0.861 & 0.393 \\ 

     \cline{2-3} 
    & \multirow{2}{*}{\makecell[c]{\rotatebox{90}{fcAR}}}  & UL &  \textbf{91.770} &  52.901 &  1.079  & 0.822 & 0.443 \\ 
    
    \cline{3-3}  
    &  & SL &  91.713 &  \textbf{85.915} &  \textbf{1.049}  & \textbf{0.802} & \textbf{0.458}  \\ 
    
		\specialrule{0em}{0pt}{0em}
		\bottomrule[1pt]
	\end{tabular}
	\label{tab:peformance}
\end{table}

\vspace{-0.2cm}
\textbf{Comparison with other models.}
Table~\ref{tab:models} presents the results of various models on the DeLTA test set. The DNN composed of fully connected multilayer perceptrons has the simplest structure and the worst performance. 
The result of CNN is better than that of DNN, which illustrates the effectiveness of the convolutional layer for feature extraction. 
It is worth noting that the CNN-Transformer, which incorporates one  Transformer \cite{Transformer} encoder with multi-head attention and residual structure, achieves better results on the \textit{F-score} \cite{metrics} of the balance of precision and recall, but for AEC accuracy, it is inferior to CNN.
The reason may be that the dataset used in this paper is not large enough, and CNN-Transformer is overfitting with the training set, resulting in poor performance on the test set. 
Previous work \cite{survey} also shows that Transformer-based models generally outperform CNN-based models on large-scale datasets, but not on small datasets. 
Furthermore, Table~\ref{tab:models} presents the results of PANNs, which achieve state-of-the-art CNN-based performance on AudioSet.
 Since the model in this paper performs both AEC and ARP tasks, we replace the last layer of PANNs with a parallel AEC layer and ARP layer, where the AEC layer contains 24 units with the sigmoid activation function, and the ARP layer contains 1 unit with the linear activation function.
 The results of PANNs in a fixed mode illustrate the importance of convolutional weights containing 527 classes of audio event knowledge for the AEC task.
In fine-tuning and fixed modes, PANNs with large-scale audio event knowledge from AudioSet are highly accurate for AEC but poor for ARP. 
Ultimately, the proposed hierarchical graph constructed with the AE information achieves competitive results in both AEC and ARP tasks.

\begin{table}[b] \footnotesize 
\setlength{\abovecaptionskip}{0.1cm}   
	\setlength{\belowcaptionskip}{-0.2cm}
	\renewcommand\tabcolsep{1pt} 
	\centering
	\caption{Comparison of different models on DeLTA dataset.}
	\begin{tabular}{
	p{2.6cm}<{\centering}| 
	p{1.3cm}<{\centering}
 p{1cm}<{\centering}|
 p{0.9cm}<{\centering}
 p{0.9cm}<{\centering}
 p{0.9cm}<{\centering}
	} 
	    \hline 
		\multirow{2}{*}{\makecell[c]{Model}} &  
\multicolumn{2}{c|}{ AEC} & \multicolumn{3}{c}{ ARP} \\

    \cline{2-6}
  &  \textit{F-score}(\%) & \textit{Acc.}(\%) & \textit{MSE} & \textit{MAE} & \textit{R2}  \\
  
		\hline
		DNN  &  53.986 & 90.049 &  1.733 & 1.011 & 0.105 \\ 

  CNN  & 55.050 &  90.750 & 1.675 & 0.997 & 0.135 \\

  CNN-Transformer  & 58.667  &  88.942 & 1.445 & 0.966 & 0.254  \\

  PANNs (Fixed)  &  53.753 &  91.058 & 1.262 & 0.880 & 0.348 \\

 PANNs (Fine-tuning)  & 63.860 &  \textbf{91.882} & 1.162  & 0.858 & 0.400 \\

 HGRL-fAR  & 67.428 & 91.751 & 1.176 & 0.861 & 0.393 \\

 HGRL-fcAR-UL  & \textbf{68.269} &  91.770 &  1.079 &  0.822  & 0.443 \\

 HGRL-fcAR-SL  & 67.911 &  91.713 &  \textbf{1.049} & \textbf{0.802} & \textbf{0.458} \\
	 
	\hline
	\end{tabular}
	\label{tab:models}
\end{table}

\begin{figure}[t] 
	\setlength{\abovecaptionskip}{0.1cm}  
	\setlength{\belowcaptionskip}{-0.6cm}   
	\centerline{\includegraphics[width = 0.5 \textwidth]{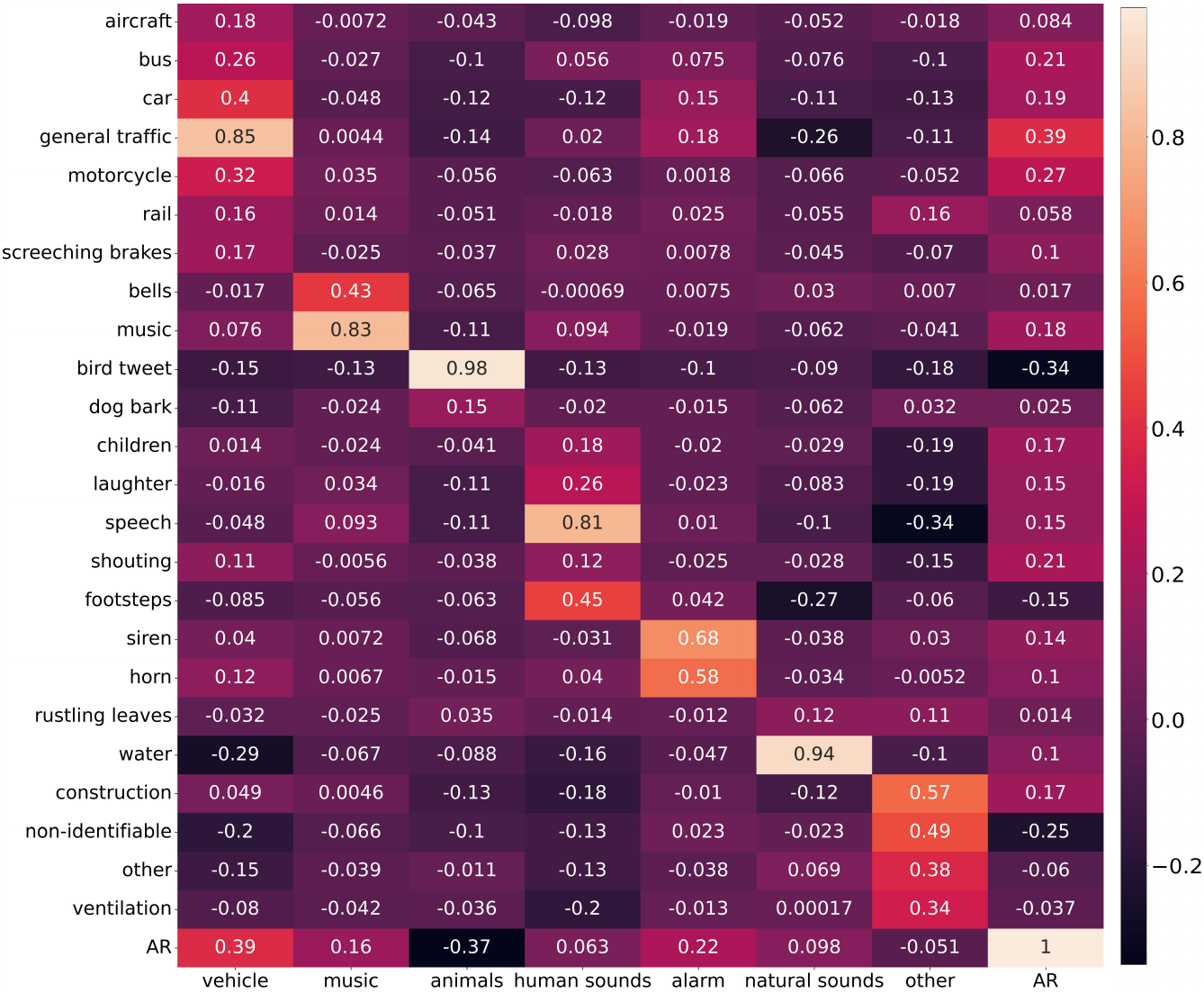}}
	\caption{Correlations of outputs by HGRL on the test set.}
	\label{event_relation}
\end{figure}

\textbf{ In-depth analysis.}
In the fcAR graph, it is worth exploring whether the model learns the relations between the introduced 7-class cAE, the existing 24-class of fAE, and the AR with corresponding supervision information.
To answer this question, Figure~\ref{event_relation} visualises the Pearson correlation coefficient (PCC) \cite{pcc} of AE probabilities and corresponding AR outputs on the test set by the fcAR-SL instead of fcAR-UL. The prediction about cAE in fcAR-UL is randomised.
The results in Figure~\ref{event_relation} show that the fcAR-based HGRL automatically aligns fAE with the cAE classes to which they belong, even if the introduced cAE information in training is inaccurate and the corresponding relations are implicit. 
Furthermore, the model also well coordinated the relations between 7-class cAE and AR, and between 24-class fAE and AR. 
Among cAEs, the most annoying AEs are the \textit{vehicle} and \textit{alarm} sounds. Conversely, \textit{animals} sounds are the least likely to be annoying. Among fAEs, the most annoying AE is the \textit{general traffic} sound, and the least annoying is the \textit{bird tweet}. These trends are consistent with human intuitions in real life.
The proposed HGRL automatically learns the relations between fAE, cAE and AR implicit in the dataset based on graph representations.
In summary, the proposed HGRL successfully captures the relations between the two different grains of AE information and further aligns them with the AR information for the joint AEC and ARP tasks.

\label{ssec:figure-f}

\vspace{-0.2cm}
\section{Conclusions}
This paper presented a method for associating the audio events-related AEC task to the human mood-related ARP task, using a HGRL model consisting of fAE, cAE and AR embeddings, inspired by the human perception of AE in real-life soundscapes. 
Experiments on the DeLTA dataset show that: 1) The HGRL using AE and AR information can effectively combine these 2 types of information and improve the performance on both AEC and ARP tasks.
2) Compared to the fAR-based HGRL, the fcAR-based HGRL using additional cAE information and the pretrained convolutional weights achieves better results on both AEC and ARP tasks.
3) The correlation-based analysis shows that the HGRL successfully integrates the AE information with the AR information for the joint AEC and ARP tasks, while capturing the relations between two different grains of cAE and fAE information implied in the dataset and further aligning them with the AR information.

 \vspace{-0.2cm}
 \section{ACKNOWLEDGEMENTS}
\label{sec:ACKNOWLEDGEMENTS}
The WAVES and AIRO Research Groups received funding from the Flemish Government under the “Onderzoeksprogramma  \quad  Artificiële Intelligentie (AI) Vlaanderen” programme.

\bibliographystyle{IEEEtran}
\bibliography{template}

\end{document}